# What makes a gesture a gesture? Neural signatures involved in gesture recognition

Maria E. Cabrera, Keisha Novak, Daniel Foti, Richard Voyles, Juan P. Wachs

*Abstract*— Previous work in the area of gesture production, has made the assumption that machines can replicate "human-like" gestures by connecting a bounded set of salient points in the motion trajectory. Those inflection points were hypothesized to also display cognitive saliency. The purpose of this paper is to validate that claim using electroencephalography (EEG). That is, this paper attempts to find neural signatures of gestures (also referred as *placeholders*) in human cognition, which facilitate the understanding, learning and repetition of gestures. Further, it is discussed whether there is a direct mapping between the placeholders and kinematic salient points in the gesture trajectories. These are expressed as relationships between inflection points in the gestures' trajectories with oscillatory mu rhythms (8-12 Hz) in the EEG. This is achieved by correlating fluctuations in mu power during gesture observation with salient motion points found for each gesture. Peaks in the EEG signal at central electrodes (motor cortex; C3/Cz/C4) and occipital electrodes (visual cortex; O3/Oz/O4) were used to isolate the salient events within each gesture. We found that a linear model predicting mu peaks from motion inflections fits the data well. Increases in EEG power were detected 380 and 500ms after inflection points at occipital and central electrodes, respectively. These results suggest that coordinated activity in visual and motor cortices is sensitive to motion trajectories during gesture observation, and it is consistent with the proposal that inflection points operate as *placeholders* in gesture recognition.

## I. INTRODUCTION

Gesturing is a form of engaging the body into expressions with varied objectives, among which are: conveying a message, completing an action, or as a reflection of a bodily state. There has been abundant research relating gestures to speech on the neurological level [1],[2], as well as the importance and benefits of gestures in learning [3]–[6].

However, this line of research has not crossed over to fields such as gesture recognition. The state of the art in gesture recognition heavily relies on data mining and visual characteristics [7],[8], yet the cognitive processes related with gesture production and perception have not been considered as a prominent source of features for gesture recognition.

By observing how humans learn to gesture and what determines the forms of gestures produced, we can learn how to generate "human-like" gestures by machines. Furthermore, if there are cognitive signatures related to gestures observed, those signatures may be used to "compress" a gesture in our memory while keeping the intrinsic characteristics of the gesture. When a gesture is recalled, these key-points associated with the cognitive signatures are used to "uncompress" these gesture into physical expression. In addition, multiple instances of the same gesture will share key common motion components among all instances, regardless of the variability associated with human performance. The purpose of this research is to find a relationship between the salient points in gesture motion with oscillations in the mu frequency band, brain activity associated with the motor cortex.

Relative to a pre-stimulus baseline, mu activity is suppressed when individuals perform gestures themselves and when they observe others perform gestures [9], [10], a phenomenon known as event-related desynchronization. Critically, this mu phenomenon is continuous throughout stimulus duration when the gestures are meaningless but follows a more dynamic time course when the gestures are social or communicative in nature [11]. Fluctuations in mu activity while observing communicative gestures is not related to the total amount of movement per se; rather, it is thought to relate to the processing of gesture meaning. Building on this finding, we tested whether the timing of mu oscillations during gesture observation would follow a unique time course for each gesture; those peaks in mu activity could be used to infer the time that most salient gesture characteristics occur. We hypothesized that these characteristics would correspond to inflection points in the kinematic trace of the gesture manifestation.

The potential impact of this finding is that we could use these inflection points to represent placeholders to produce human-like instances of gestures non-discriminable by the human psyche.

The contributions of this paper are: (a) quantify the unique time courses of oscillatory mu activity elicited by a range of communicative gestures; (b) establish a link between the timing of mu oscillations and the occurrence of salient points in gesture motion trajectories; and (c) demonstrate that inflection points in gesture motion elicit corresponding peaks in mu activity, suggesting that they may operate as placeholders in gesture recognition.

The remaining sections of this work cover a background on the importance of gestures as utterance and the neurological aspects of action understanding. This is followed by the materials and methods section, where the experiment design is described. The results section presents the main findings, complemented by a discussion section. Finally, the conclusions and recommendations are presented.

## II. BACKGROUND

### A. Gesture Utterance

While different gesture representations have been developed and are well established in the machine learning community, there is still a lack of research that includes the cognitive/behavioral aspects and the computational aspects in a coherent schema that can be applied to gesture recognition.

Reviewed studies show analysis on gestures from different frameworks: Novack et al. [12] analyzed gestures from a movement perspective regarding surrounding objects and the impact on recognizing motions as gestures or not. Gibet et al. [13] analyzed gestures from a motor control perspective finding some commonalities in gesture execution related to invariance in velocity profiles and minimum jerk. This supports the idea that gestures are executed through smooth trajectories.

Regarding gesture perception, Peigneux et al. [14] conducted a study using brain imagery to detect the areas of the brain involved in the imitation of familiar and novel gestures, with the aim to build a cognitive model for apraxia (inability to perform particular purposive actions, as a result of brain damage). Their findings suggest that a single memory store may be used for gestural representation either in perception or production.

Brain functionality related to both gestures and spoken language appears to be highly overlapped. This conveys the importance of gestures in both adult speech and infants' language development [15]. Kok and Cienki [16] highlight the similarities in the cognitive process relating speech with gestures, aligning some of the properties of cognitive grammar to gesture production. A study based on preschool children [17] compared the effects of including physical activity and gesturing when learning a foreign language. The use of task-relevant gestures had a positive effect on learning, showing consistent results with previous research relating gestures with lasting memories [18].

### B. Action Understanding and the Brain

Understanding the actions of others is critical to assess goals and intentions. This fundamental recognition is essential for survival, as well as social interactions and functioning. In terms of survival, and animal's ability to identify movements of potential mates, prey, and predators are essential for predicting future movements where the consequences of such movement may prove disastrous. Similarly, humans require such fine-tuned perception in order to function at a much higher level. Specifically, empathy, communication, social skills, and overall coordination can all be traced back to the importance of motor representations [19].

Mirror neurons in the premotor and parietal cortex of monkeys have been shown to become activated when the monkey performs a particular action, and when it observes others behaving in similar ways [20]. The mirror neuron system (MNS) appears to be activated by all actions, both simple and complex, and also during the execution as well as observation of an action [19].

There have also been claims that MNS activation is closely connected to electrophysiological mu suppression [21]. Mu suppression can be measured non-invasively by electrodes on the scalp. As stated previously, mu suppression has been argued to play a role in contexts involving social interaction as well as passive action observation. However, when gesticulations are meaningful, representations of this semantic meaning should also be mapped to the visual stimuli in order to determine if there is a match between meaning and the representation of an action.

## III. MATERIALS AND METHODS

The data used in this work was gathered and analyzed in two distinct fronts represented in Fig. 1: one is related to the motions by the person performing the gestures, where the trajectories of the hands are tracked, and used to determine the occurrence of inflection points; the second is related to the neural signatures acquired from the person observing the performed gestures. These neural signatures are analyzed to find the occurrence of signal peaks in the EEG channels related with the motor and visual cortex.

Motion information was extracted from each gesture in one lexicon from the ChaLearn dataset. For each gesture, time stamps for inflection point occurrences were compiled. Inflection points were determined taking the first derivative of the motion trajectory for each of the gesturer's hands, and finding the local maxima and their corresponding time stamp in the gesture duration.

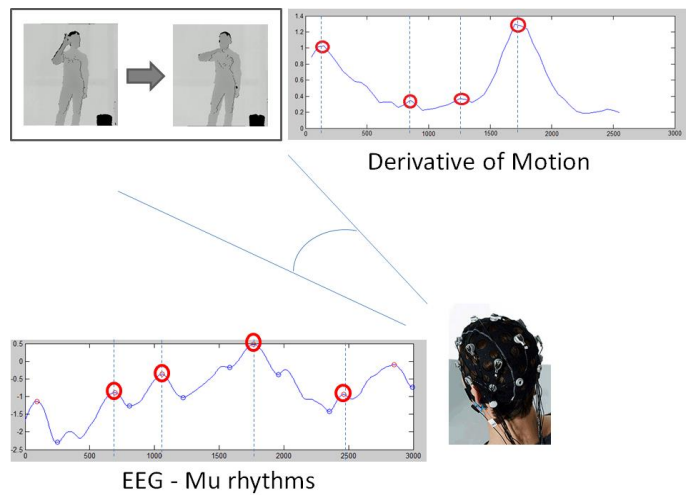

Fig. 1. Methodology Overview

### A. Participants

The data presented was collected from four adult volunteers, as part of a larger ongoing project (Age: M=20.5, SD=3.0, Range=18-24; Gender: 1 female, 3 male).

### B. Gesture Task

Participants passively viewed 20 videos taken from the 2013 ChaLearn stimulus set [22]. Each video was approximately 3 seconds in length and showed a single actor performing a gesture. Videos were presented in black and white and only the silhouette of the actor was visible, thereby making the physical motion of the gesture the most salient aspect of the video. Participants viewed each gesture a total of six times, with an inter-stimulus interval of 2 seconds; gesture order was randomized across participants.

*C. EEG and Analysis*

The continuous EEG was recorded using an ActiCap and the ActiCHamp amplifier system (Brain Products). The EEG signal was digitized at 24-bit resolution and a sampling rate of 500 Hz. Recordings were taken from 32 scalp electrodes based on the 10/20 system, with a ground electrode at Fpz. BrainVision Analyzer (Brain Products) was used for offline analysis. Data were referenced to the average mastoid electrode and bandpass filtered from 0.1-100 Hz. The signal was segmented from -1000 to 5000ms relative to video onset, and ocular correction was performed using a regression algorithm. Artifacts were rejected within individual channels using a semiautomated procedure. Oscillations in the mu frequency band were evaluated using a continuous wavelet transformation. Complex Morlet wavelets were applied, with a frequency range of 1-20 Hz, frequency steps of 0.5 Hz, and Morlet parameter of c=5. Segments were baseline corrected relative to the -500 to -100ms window and averaged separately for each gesture video. We focused our analysis on the wavelet layer peaking at 10 Hz. We extracted this wavelet at a cluster of central electrodes (i.e., directly above the motor cortex; C3/Cz/C4), where mu activity is known to be maximal. For comparison, we also considered an occipital cluster (O3/Oz/O4). We expected that these occipital electrodes would capture visual cortical activation associated with stimulus processing but not mu activity per se [11].

*D. Statistical Analysis*

For this analysis, the three channels associated with the motor cortex responses (C3/Cz/C4) were averaged over and will be further referred to as C Channels; analogous to the channels associated with visual cortical activation (O3/Oz/O4) further referred to as the O Channels. Inflection points extracted from one ChaLearn's lexicon, are the peaks found in the first derivative for the motion trajectory of the gesturer's hands for each gesture class.

Time stamps of inflection points from motion data and peak information from EEG data were plotted (time vs. time). Three different trend lines were fitted and their corresponding $R^2$ values were extracted to determine correlation between the variables.

A linear regression model was fitted for three different scenarios:

- The dependent variable was the time response for the C Channels, when the independent variables considered are the motion data and the gesture class, including the interaction between them. This determines a relationship between responses in the motor cortex when there are salient points in gesture motion.

- The dependent variable was the time response for the O Channels, when the independent variables considered are the motion data and the gesture class, including the interaction between them. This determines a relationship between responses in the visual cortex when there are salient points in gesture motion.

- The dependent variable was the time response for the C Channels, when the independent variable considered is the O Channels. This determines a relationship between activation in the motor cortex when there are activations in the visual cortex

Statistical significance of the linear regression coefficients was determined in each case using F-test and p-values for each independent variable.

IV. RESULTS

*A. EEG Response*

First, the average EEG response was collapsed across all gesture stimuli in order to test for commonalities in mu activity. At occipital electrodes, there was a decrease EEG power at 10 Hz, which corresponds to alpha activity and represents visual cortex activation associated with stimulus viewing. As expected, this average response at occipital electrodes was sustained from 400-3000ms, indicating continuous visual activity throughout stimulus duration and that participants attended to the gestures videos.

A different pattern emerged at central electrodes, where mu activity occurs: there was evidence of mu suppression from 180-1000ms and peaking at 400ms (highlighted by the gray box in Fig. 2, top). This indicates that, regardless of the unique kinematic characteristics of each gesture, there was automatic suppression of mu activity within the first second of gesture viewing. This is consistent with a previous EEG study showing that both communicative and meaningless gestures elicited common mu suppression during the first second of viewing, a general response within the motor cortex that is distinct from the visual cortical response apparent at occipital electrodes [11].

Next, we considered the latencies of the first positive-going peak of mu activity for each gesture (i.e., the first positive local maximum occurring after stimulus onset). Insofar as each gesture video elicits a decrease in mu power within the first 400ms of stimulus viewing, these positive-going peaks represent the first evidence of dynamic mu oscillatory activity associated with gesture interpretation. Of the 20 gestures used, 14 (70%) elicited a peak in mu activity within 1000ms of the inflection point for that video; for the remaining 4 gestures, the mu peak either occurred before the inflection point (3) or more than 1000ms afterward (1), indicating that it was likely not associated with the inflection point per se. Of these 14 gestures, the time interval from the inflection in the gesture video to the peak in mu activity was 272ms (SD=163ms, Range=14-632ms). This pattern indicates correspondence between the timing of mu oscillations and initial gesture inflection points.

Of the 20 gesture stimuli, 16 contained a second identifiable inflection point. Of these, 9 (56%) also elicited a second peak in *mu* activity; the remaining 7 gestures either elicited a peak before the second inflection point (6) or did not elicit a second *mu* peak (1). Of these 9 gestures that had a second inflection point and a subsequent second peak in *mu* activity, the pattern was similar to that of the first peak, with an average inflection-peak interval of 339ms (*SD*=228, Range=8-654ms).

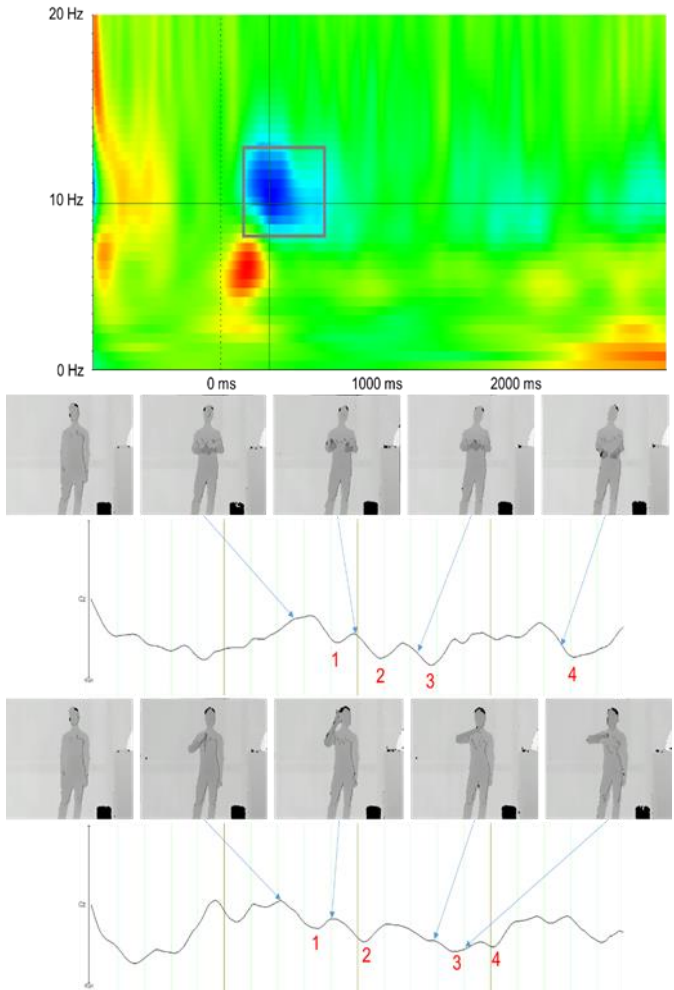

Fig. 2. Top: EEG power, averaged across all gesture stimuli (x-axis=time, y-axis=frequency; blue=decrease in power, red=increase). Bottom: Time course of mu activity (10 Hz wavelet) for two representative gestures, each of which had four inflection points, shown in the embedded video frames. Arrows show the timing of each inflection point, and the corresponding mu peaks are labelled in red. Note that positive power is plotted down in the two mu graphs.

Of particular interest were two gesture stimuli that each had four identifiable inflection points and also elicited four mu peaks. The time courses of mu activity, as well as the relative positioning of the inflection points and the mu peaks for each gesture, are presented in Figure 2 (bottom). For both of these gestures, there was a relatively consistent inflection-peak lag, with all for peaks occurring within 300ms of each inflection point (Gesture #7: *M*=163ms, Gesture #12: *M*=244ms). Thus, for gestures with relatively rich kinematic characteristics (i.e., multiple inflection points occurring within 3s), there appears to be close correspondence with the timing of these inflection points and the timing of fluctuations in mu activity. These preliminary results are consistent with the proposition that inflection points function as placeholders within gesture expression, and that these placeholders modulate activity within the motor cortex that is associated with gesture processing and interpretation.

### B. Statistical Analysis

Collected timestamps in both motion trajectory and EEG channels were used to determine a correlation between the variables, namely inflection points occurrence and peaks in mu activity. A time vs. time plot was generated to find the linear relationship between responses in three different combinations: motor cortex responses in terms of inflection point occurrence (C vs. IP, Fig. 3), visual cortical activation in terms of inflection point occurrence (O vs. IP, Fig. 4), and motor cortex responses in terms of visual cortical activation (C vs. O, Fig. 5). These plots were generated without regarding gesture class and focusing on the time responses in the lexicon set.

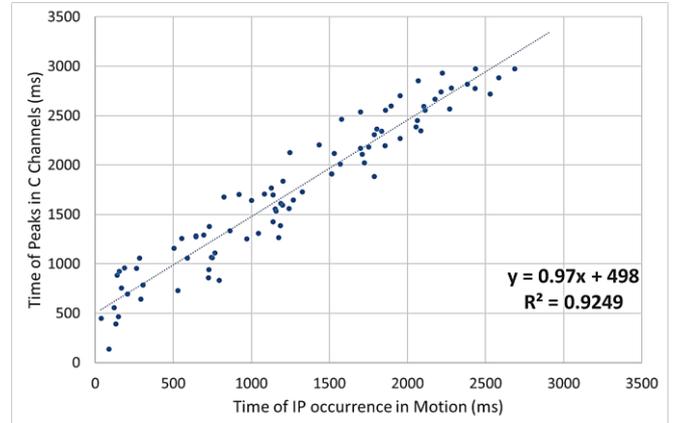

Fig. 3. Time points of motor cortex responses in terms of inflection point occurrence (C vs. IP). Fitted linear trendline with $R^2 = 0.925$

For the first combination, C Channels vs. inflection points, the fitted linear equation, with coefficient of determination $R^2 = 0.925$, shows an independent term associated to the lag between occurrence of an inflection point and a peak in the motor cortex response of 498ms.

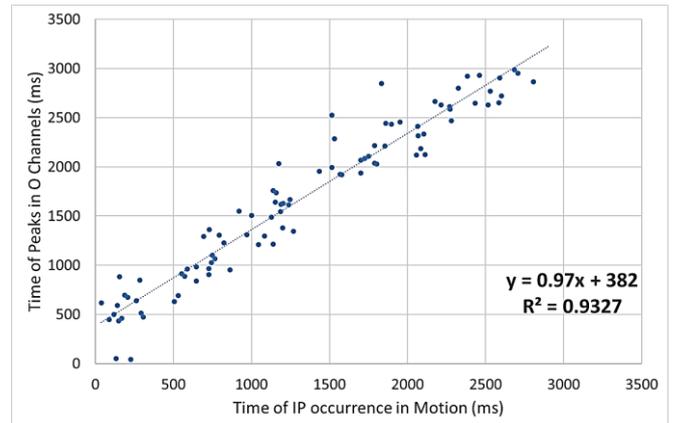

Fig. 4. Time points of visual cortical activation in terms of inflection point occurrence (O vs. IP). Fitted linear trendline with $R^2 = 0.933$

For the second combination, O channels vs. inflection points, the fitted linear equation with $R^2 = 0.933$, shows an independent term of 382ms. This once again represents the lag between the occurrences in the two signals.

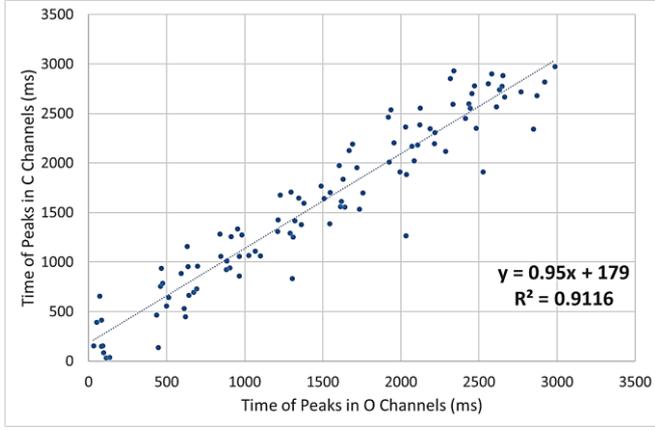

Fig. 5. Time points of motor cortex responses in terms of visual cortical activation (C vs. O). Fitted linear trendline with $R^2 = 0.912$

The last combination, motor cortex responses in terms of visual cortical activation, resulted in a fitted line with $R^2 = 0.912$, the lowest coefficient among all combinations. In all combinations, positive correlation was found among the variables. In all cases, the adjusted slope value was close to 1, indicating a close to one-to-one correspondence among responses.

A second analysis was conducted on the data, this time using the information of each gesture class. A linear regression model was fitted for two of the previous combinations, using an additional independent variable which represented gesture class and its interaction with time responses of inflection point occurrence.

The model used for the first combination is shown in (1). The fitted model showed an F-statistic of 357 with p <0.001.

$$C_{Ch} \sim 1 + G_{Class} * M_{IP} \qquad (1)$$

None of the coefficients associated with gesture class showed statistical significance, while the coefficients for the intercept (estimated 187.4) and the time occurrences for inflection points from the motion data did (p<0.0001).

Similar results were found for the second fitted model shown in (2), with an F-statistic of 410 with p<0.001. In this case the intercept estimation was 165.8 (p<0.0001).

$$O_{Ch} \sim 1 + G_{Class} * M_{IP} \qquad (2)$$

## V. Discussion

Consistent with previous research, gesture observation was associated with mu suppression during gesture observation. Whereas previous studies have focused on static measures of mu suppression (i.e., average reduction in EEG power from during the full stimulus presentation), here we focused on the time course of oscillatory mu activity and whether the timing of peaks in mu power would correspond to the occurrence of salient gesture characteristics. We found a strong linear relationship between the timing of inflection points in the gesture motions and mu peaks within the EEG signal, with an average lag of approximately 500ms. In other words, we found close correspondence with the time course of gesture motion and the time course of mu oscillations. This result is consistent with the notion that inflection points operate as placeholders involved in gesture recognition.

Notably, we detected a similar pattern of oscillatory activity at occipital electrodes. Mu activity, thought to represent activation of the motor cortex during both action execution and action observation, is maximal at central electrodes (i.e., those that are directly above the motor cortex [9]. On the other hand, EEG activity in this same frequency band, 8-12 Hz, at occipital electrodes is known as alpha and is thought to represent activation of the visual cortex. In a previous study, observation of communicative versus meaningless gestures modulated central mu activity but not occipital alpha activity [11]. That study only considered average EEG power in 1s intervals, however, whereas in the current study we used the temporal precision of wavelets to detect peaks within the EEG waveform. We found a close coupling between the timing of peaks at central and occipital electrodes, and both were related to the timing of inflections in the motion trajectories. One possibility is that this represents coordinated activation of visual and motor cortices involved in gesture processing. The earlier timing of occipital peaks (380 vs. 500ms) suggests that visual cortical activity may have been driving motor cortical activity, although this needs to be verified in future research.

The current findings are qualified by several limitations. Insofar as these pilot data were collect to assess the feasibility of using EEG to detect placeholders in gesture recognition, the sample size was small; these findings require replication in a larger sample. We also only considered communicative gestures. In a future study, it would be valuable to contrast communicative versus meaningless gestures in order to test whether visual-motor EEG oscillations are specific to gesture recognition (i.e., communicative gestures only) or generalize to all instances of biological motion, regardless of communicative intent.

The current findings demonstrate that oscillations in EEG mu rhythms are sensitive to the occurrence of inflection points during gesture processing. This is consistent with the possibility that these inflection points are salient characteristics involved in gesture recognition, comprehension, and repetition.

## VI. Conclusion

The purpose of this paper is to validate the claim that gesture production contains a bounded set of salient points within the motion trajectory by finding a relationship between these salient points in motion trajectory with the neural signatures of gestures using EEG. This was achieved by correlating fluctuations in mu power during gesture observation with salient motion points found for each gesture. Peaks in the EEG signal at central electrodes (motor cortex; C3/Cz/C4) and occipital electrodes (visual cortex; O3/Oz/O4) were used to isolate the salient events within each gesture. The quantified time courses for mu activity oscillation were detected 380 and 500ms after inflection points at occipital and central electrodes, respectively. These results suggest that coordinated activity in visual and motor cortices is sensitive to motion trajectories during gesture observation, and it is consistent with the proposal that inflection points operate as *placeholders* in gesture recognition.

The potential of this work is that it provides evidence that inflection points are "placeholders" for key points in the gesture trajectory, which "encode" the gestures in human cognition. Therefore, these points can be used to capture large variability within each gesture while keeping the main traits of the gesture class. The limitations of these work include the lexicon size and the sample size of participants for gathered data. Further studies with other type of gestures and larger lexicons should be conducted. Future work should include understanding how these placeholders appear in different level of abstractions in gestures. For example, the study focused on trajectories of the hands, but does not consider the motion of the fingers or the hand or body configuration. We believe that similar inflections points are likely to exist in the spatial-temporal kinematic traces of gestures in the other abstraction levels as well, but this remains to be tested in future work.

## *Acknowledgment*

Acknowledgements will be included after the anonymized review process is completed.